# Dark X-ray Galaxies in the Abell 1367 Galaxy Cluster

Mark J. Henriksen[1], Scott Dusek[1]

**Abstract:**

We have characterized a sample of extended X-ray sources in the Abell 1367 galaxy cluster that lack optical counterparts. The sources are galaxy size and have an average total mass of $1.3 \times 10^{11}$ Solar Masses. The average hot gas mass is $3.0 \times 10^9$ Solar masses and the average X-ray luminosity is $4.3 \times 10^{41}$ ergs cm$^{-2}$ s$^{-1}$. Analysis of a composite source spectrum indicates the X-ray emission is thermal, with temperature of 1.25 – 1.45 keV and has low metallicity, 0.026 – 0.067 Solar. The average hot gas radius (12.7 kpc) is well matched to nominal stripping radius. We argue that this optically dark, X-ray bright galaxy (DXG) population forms by a sequence of stripping followed by heating and mixing with the intracluster medium.

Corresponding author henrikse@umbc.edu

[1] Department of Physics, University of Maryland, Baltimore County, 1000 Hilltop Circle, Baltimore MD 21250

## 1. Introduction

Forty-seven ultra-diffuse galaxies (UDGs) were found in the Coma cluster (van Dokkum et al. 2015).The galaxies are generally Milky Way size but have very low surface brightness as a result of their large size combined with little star formation. The stellar masses are characteristically small, ~$10^8$ Solar masses. The number of UDGs was substantially increased to 854 in a later study (Koda et al 2015). This extensive study showed that the large sample of UDGs is concentrated around the cluster center. This, together with a low baryon fraction, suggests that the galaxies experienced an episode of gas removal in the past. They have a passively evolving red population indicating that star formation was subsequently quenched. One galaxy in particular, Dragonfly 44, has been studied extensively using its unexspectedly large globular cluster population in a dynamical study. This study gives an extrapolated mass of ~$10^{12}$ Solar masses (van Dokkum et al. 2016), consistent with a very low baryon fraction, < 1%, that characterizes UDGs (Koda et al. 2015). Use of weak lensing to determine the total mass, for a large sample of UDGs within nearby galaxy clusters, has further defined the mass characteristics: a low baryon fraction and a mean total mass of ~$10^{11}$ Solar masses within < 30 kpc (Sifon et al. 2018).

In contrast, study of UDGs in the Hydra I cluster finds lower characteristic masses, < $10^{10}$ Solar masses, few globular clusters (GC), and a factor of 10 lower stellar mass: in the range of $10^7$ - $10^8$ Solar masses. This suggests that Dragonfly 44 may be unique in its large mass and GC content (Iodice et al. 2020). UDGs are also found to significantly populate clusters in the redshift range: 0.31 < Z < 0.55, at numbers that imply they form during cluster evolution

(Janssens et al. 2019). UDGs are also not evenly distributed within the cluster. Their projected spatial distribution is azimuthally asymmetric, with deficiencies in the regions of highest mass density (Janssens et al. 2019).

One explanation for UDGs occurring in the cluster environment is that galaxies with mass range $10^{10} - 10^{11}$ Solar masses, have little star formation due to environmental effects. An episode of interstellar gas removal followed by heating is suggested as the cause of star formation being quenched in these galaxies (Carleton et al. 2019). A detailed study of the nearby UDG, UGC 2162 (Trujillo et al. 2017), gives a virial mass of $10^{11}$ Solar masses. The galaxy has a low star formation rate, small stellar population, and relatively large ($10^8$ solar mass) HI gas reservoir. This galaxy may be a precursor to the UDG morphology. As these authors (Trujillo et al. 2017) point out, truncation of star formation followed by passive evolution of UGC 2162 would eventually make the galaxy, faint, red, and perhaps undetectable optically. In this paper we present a new endpoint for such galaxies, visibility in the X-ray. Abell 1367 shows signs of a merger (Ge et al. 2019) (Cortese et al. 2004), and has examples of galaxies undergoing ram-pressure stripping (Consolandi et al. 2017) making it an ideal cluster to study galaxy evolution in the X-ray.

## 2. X-ray Data

Table 1: Parameters from the 3XMM DR8 Catalogue

| RA | | | Dec | | | Flux | Flux Error | Posn error | Rate | Error | Extent | Extent Error |
|---|---|---|---|---|---|---|---|---|---|---|---|---|
| hour | min | sec | deg | arcmin | arcsec | ergs/cm²/s | | arcsec | cps | | arcsec | |
| 11 | 44 | 44.03 | 19 | 42 | 13.5 | 3.79E-13 | 8.61E-14 | 0.9 | 2.1E-01 | 2.5E-02 | 72.13 (33 kpc) | 3.8 |
| 11 | 44 | 57.99 | 19 | 40 | 19 | 1.14E-13 | 5.83E-14 | 4.1 | 6.7E-02 | 9.8E-03 | 27.92 (12.8 kpc) | 2.6 |
| 11 | 45 | 4.702 | 19 | 40 | 40.4 | 3.49E-13 | 9.42E-14 | 1.1 | 2.0E-01 | 3.4E-02 | 67.26 (31 kpc) | 3.7 |
| 11 | 45 | 6.274 | 19 | 36 | 9.3 | 4.77E-13 | 2.75E-13 | 5.8 | 1.0E-01 | 2.9E-02 | 24.72 (11.4 kpc) | 3.4 |
| 11 | 44 | 40.81 | 19 | 50 | 38.1 | 3.06E-14 | 9.47E-15 | 5.8 | 1.6E-02 | 3.7E-03 | 17.47 (8.0 kpc) | 3.2 |
| 11 | 44 | 13.15 | 19 | 59 | 17.2 | 7.04E-14 | 2.42E-14 | 2.8 | 2.1E-02 | 4.6E-03 | 92.65 (4.3 kpc) | 2.2 |
| 11 | 43 | 48 | 19 | 58 | 22.7 | 1.82E-13 | 3.05E-14 | 1.8 | 8.2E-02 | 5.8E-03 | 16.34 (7.6 kpc) | 1.1 |
| 11 | 43 | 9.843 | 19 | 42 | 54 | 5.36E-15 | 9.16E-14 | 5.8 | 4.5E-03 | 6.6E-03 | 27.02 (12.4 kpc) | 3.9 |
| 11 | 43 | 43.53 | 19 | 59 | 7.46 | 5.10E-14 | 3.57E-14 | 5.0 | 1.7E-02 | 3.9E-03 | 20.53 (9.44 kpc) | 2.6 |
| 11 | 43 | 42.9 | 19 | 59 | 14.6 | 1.82E-14 | 9.12E-15 | 3.0 | 1.2E-02 | 2.2E-03 | 8.8 (4.0 kpc) | 1.5 |
| 11 | 43 | 39.63 | 20 | 0 | 14.9 | 3.83E-14 | 1.62E-14 | 5.4 | 2.2E-02 | 5.7E-03 | 23.75 (10.8 kpc) | 2.8 |
| 11 | 43 | 35.32 | 20 | 0 | 13.7 | 3.13E-14 | 1.42E-14 | 2.3 | 9.3E-03 | 1.7E-03 | 6.70 (3.1 kpc) | 1.4 |
| 11 | 42 | 58.52 | 19 | 44 | 32.2 | 1.059E-13 | 5.21E-13 | 2.3 | 6.7E-02 | 3.8E-02 | 42.90 (19.7 kpc) | 8.2 |
| 11 | 43 | 30.13 | 20 | 3 | 40.2 | 1.20E-14 | 1.44E-14 | 4.2 | 7.9E-03 | 2.9E-03 | 10.45 (4.8 kpc) | 2.1 |

| | | | | | | | | | | | |
|---|---|---|---|---|---|---|---|---|---|---|---|
| 11 | 43 | 11.53 | 20 | 0 | 33 | 3.69E-14 | 1.95E-14 | 4.0 | 1.8E-02 | 3.9E-03 | 21.92 (10.1 kpc) | 2.7 |
| 11 | 42 | 57.34 | 19 | 59 | 34.3 | 7.75E-14 | 4.15E-14 | 4.5 | 3.7E-02 | 9.0E-03 | 43.86 (20.2 kpc) | 5.6 |

The sources we study are a subset of the third XMM-Newton serendipitous source catalog (Rosen, et al. 2016). Sixteen sources meet the following criteria: (1) extended sources in the 3XMM DR8 catalogue, (2) within 1 Mpc (36') of the optical center of the Abell 1367 galaxy, and (3) no optical counterpart. A search radius of 1 Mpc is chosen because this is the radial extent of the intracluster gas (Ge et al. 2019) , (Cortese et al. 2004) ). The 3XMM DR8 catalogue data uses the Science Analysis System (SAS) tool, *emldetect,* for source detection. Each source has a radial extent, with error. The determination of extent is made by convolving a beta-model profile (Cavaliere & Fusco-Femiano 1978) with the EPIC PN point spread function and performing a maximum likelihood fit to the source image. A size below 6 arc sec is considered to be a point source. To avoid non-convergence in the fitting procedure, an upper limit of 80" is imposed. Basic data from the catalogue, relevant to our subsequent study, is given in Table 1 for the 16 sources.

The sources are plotted on the Digitized Sky Survey (DSS) image of Abell 1367 in Figure 1, left pane, to show their general location relative to the optical light distribution in the cluster. In Figure 1, right pane, the sources are plotted on a smoothed Rosat All-Sky Survey (RASS) X-ray image of the cluster to show their location relative to the hot diffuse gas of the intracluster medium (ICM). The circles are plotted at the position of the source and their radii are the extent of the source. Both position and radius are given in Table 1.

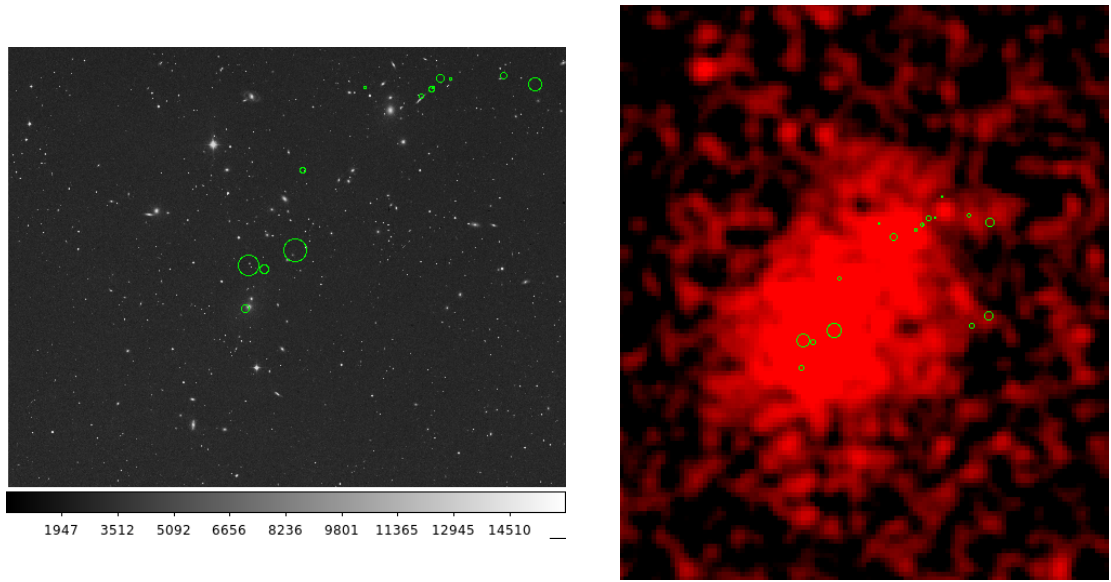

Figure 1: Extended X-ray Sources from the 3XMM DR8 Catalogue with no counterpart. Left pane is source size and position on the DSS. The right pane is source size and location on the ICM shown in the RASS image.

## 2.1 Counterpart Identifications

We used the NASA/IPAC Extragalactic Database (NED) to search for optical counterparts associated with galaxies. NED contains a comprehensive number of catalogs including the Sloan Digitized Sky Survey (SDSS), and covers the selected region of Abell 1367 very well. The SDSS is sensitive down to 25.7 mag in this region. For comparison, the Dragon Fly galaxy, in the Coma cluster, is 21st magnitude and UGC 2162 has an SDSS R magnitude of 19.63. Thus, both would be counterparts so that the lack of an optical counterpart is meaningful. The higher mass UDGs would have been identified in our cross-correlation. The $1\sigma$ positional error is used to create an error circle of radius $3\sigma$ centered on the position of the X-ray source. An extensive discussion of the positional errors, which range from 0.9 – 5.8 arcsec, is given in (Ge et al. 2019) . The 16 sources lack any optical counterpart within the error circle. Ten of the X-ray source have a galaxy partially overlapping the X-ray emission. However, the nearest galaxies are offset from 4 – 15 kpc away from the centroid of the X-ray. In section 4, we argue that the X-ray sources are not stripped remnants of these nearest galaxies. We now refer to the 16 sources as dark X-ray galaxies (DXGs) as they are dark in regards to visual light from star formation but X-ray bright.

## 2.2 Spectral Fitting

Spectral fitting is necessary to determine the nature of the X-ray emission, whether thermal or non-thermal. The sources have a continuous range of count rates which means that a composite spectrum will not be dominated by a single strong source and will truly characterize the sample. The HEASoft tool, *Xselect*, was used to extract the composite spectrum from the three XMM fields that contain the sources. Source free regions on these fields were also used to extract a background spectrum. The background subtracted spectrum was then fit using the *Xspec* tool. Two models were tested, a thermal model (Raymond and Smith, *RS*) and a power law model (*PWL*). Both models have the Galactic column density ($N_H$) and model normalization as free parameters. The thermal model also has abundance and temperature as free spectral parameters while the power law has spectral index. The residuals for each spectral fit are shown in Figure 2.

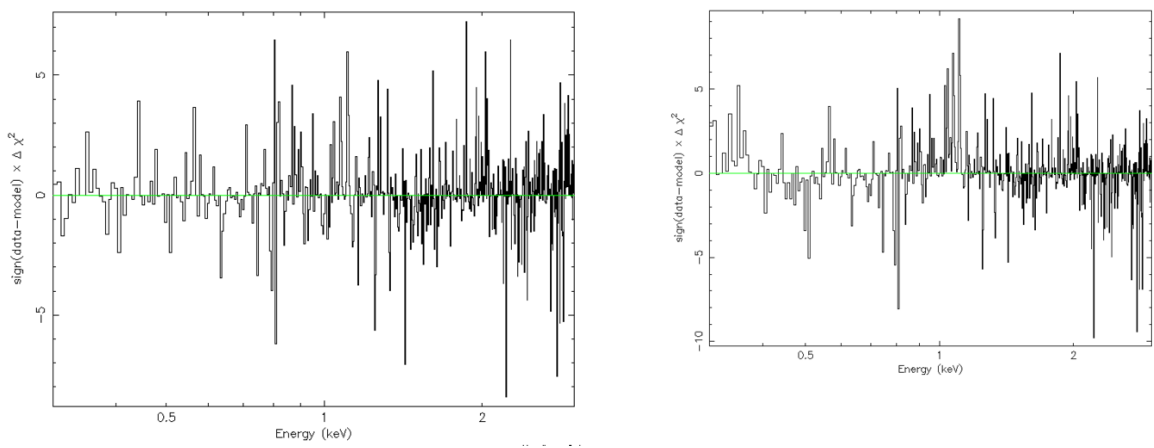

**Figure 2.** Residuals from fitting the composite spectrum of the extended X-ray sources. (**a**) The left panel shows residuals for a single component thermal model. The residuals are generally random; (**b**) The right

panel shows residuals for a power law model. There are significant residuals at low energy where the power law overpredicts the soft X-ray emission, can't model the weak line emission around 1 keV, and at greater than 2 keV where it produces too little continuum.

The 90% range on all fit parameters are given in Table 2. There are 542 PHA bins and 338 degrees of freedom (RS) and 339 (PWL). Line emission is weak for the DXGs. However, the power law model, shows some residuals around 1 keV from unmodeled weak FeL emission. The continuum has a different shape for each model and the power law gives too much soft X-ray emission compared to the thermal below 0.5 keV. The power law model requires twice the Galactic column density, in an effort to absorb the excess model emission. The chi-square residuals throughout show that a power law is a significantly worse fit to the data. Xspec calculates the null hypothesis; defined in Xspec as the probability of the observed data being drawn from the model given the minimum chi-square and the number of degrees of freedom. The probability of the thermal model being a good fit is 61%, compared to 12% for the power law. Based on this statistic and the excess residuals that show the power law is a worse fit to the data, we adopt the thermal model as the description of the DXG emission. The characteristic temperature is 1.25 – 1.45keV and the abundance is 0.026 – 0.067 Solar.

Table 2. Values obtained from fitting the composite spectrum

| RS$\chi^2$ | PWL$\chi^2$ | RS $NH$ | PWL NH | kT | Abundance | Spectral Index | RS Norm | PWL Norm |
|---|---|---|---|---|---|---|---|---|
| | | $10^{22}$ cm$^{-2}$ | $10^{22}$ cm$^{-2}$ | keV | %Solar | | | |
| 320 | 361 | 0.070-0.086 | 0.13-0.15 | 1.25-1.45 | 0.026-0.067 | 2.26-2.53 | 0.94-1.10 | 0.22-0.24 |

A temperature estimate for individual DXGs is calculated from the hardness ratio. The hardness ratios (given in Table 3) are for the (0.2 - 2 keV)/((2 - 12.0 keV) range. Using the Xspec *fakeit* tool, model hardness ratios are made using the XMM response matrix. For a range of temperatures, the ratios and their corresponding temperatures are: 2678 (0.5 keV), 28.1 (1 keV), 10.9 (1.5 keV), and 0.89 (2 keV). The range of observed hardness ratios is 2.12 - 918 with 2/3 between 2 and 10. Interpolating between these hardness ratios to find the estimated temperatures shows that they range from 0.8 - 1.9 keV. This overlaps the range of temperature found from fitting the composite spectrum: 1.25 - 1.45 keV.

Table 3: Luminosities and Temperatures

| Galaxy ID | Luminosity | Hardness Ratio | Temperature |
|---|---|---|---|
| | ($10^{41}$ ergs/s) | (0.2-2 keV)/(2 – 12 keV) | (keV) |
| DXG1 | 5.50 | 10.0 | 1.5 |
| DXG2 | 1.80 | 17.26 | 1.3 |
| DXG3 | 5.30 | 14.92 | 1.4 |
| DXG4 | 2.60 | 2.12 | 1.9 |
| DXG5 | 0.42 | 6.26 | 1.7 |
| DXG6 | 0.55 | 2.43 | 1.9 |
| DXG7 | 2.20 | 10.0 | 1.5 |
| DXG8 | 0.12 | - | - |
| DXG9 | 0.45 | 8.46 | 1.6 |
| DXG10 | 0.32 | 40.6 | 1.0 |
| DXG11 | 0.58 | 9.2 | 1.5 |
| DXG12 | 0.25 | 4.7 | 1.8 |
| DXG13 | 1.80 | 918 | 0.8 |
| DXG14 | 0.21 | 59.7 | 1.0 |
| **DXG15** | 0.48 | 8.25 | 1.6 |
| **DXG16** | 0.98 | 10.3 | 1.5 |

## 2.3 Luminosity

The fluxes in Table 1 are calculated from the 3XMM catalogued count rates using the HEASoft tool *PIMMS*. A thermal spectrum and the XMM EPIC response matrix are used for the count rate to flux conversion. The parameters are: the best fit temperature from the composite spectrum, 0.2 solar abundance, and the mean Galactic column density, $1.84 \times 10^{20}$ cm$^{-2}$ for the Abell 1367 cluster. The column density, $n_H$, is obtained from The HEASARC $n_H$ tool (Ben Bekhti et al. 2016). The luminosities are calculated using the fluxes and distance to Abell 1367. Figure 3 shows a histogram plot of the luminosities. The range in luminosity is from $1.2 \times 10^{40}$ - $5.5 \times 10^{41}$ ergs cm$^{-2}$ s$^{-1}$. The peak is at $5.6 \times 10^{40}$ ergs cm$^{-2}$ s$^{-1}$, with a higher average luminosity of $4.3 \times 10^{41}$ ergs cm$^{-2}$ s$^{-1}$. Comparison to early-type galaxies shows that the X-ray luminosities are in the middle of the range for that class of galaxies (Babyk et al. 2018) . One possible origin of the X-ray emission in the dark galaxies is stripped halo gas (Forman et al. 1979) . We discuss this possibility in Section 4.

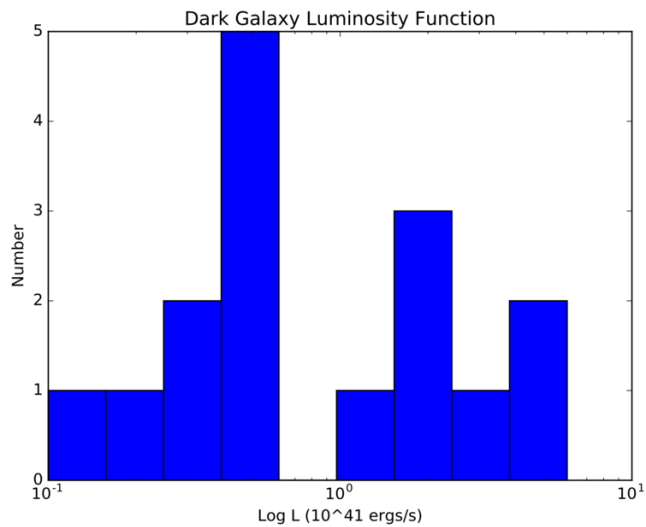

**Figure 3.** The Luminosity Histogram for Dark X Galaxies

## 3. Masses

The average ionized Hydrogen density ($n_H$), given in Table 4, is used in the gas mass calculation as any constraint on the density profiles is impossible with this data. The densities are calculated from the fluxes derived from the measured count rates for each dark galaxy. The densities scale as:

$$n_H = 9.4 \times 10^{-3} \left(\frac{L}{10^{41} ergs/sec}\right)^{1/2} / \left(\frac{R}{10 kpc}\right)^{3/2} cm^{-3} \quad (1)$$

The mass calculation assumes that the hot gas is spherically symmetric with primordial abundances, 92% Hydrogen and 8% He by number. The very low metallicity from the composite spectral fit ~0.04 Solar is consistent with the hypothesis of little star formation in an optically dark galaxy. The scaled equation for gas mass is the following:

$$M_{gas} = 1.15 \times 10^8 \left(\frac{n_H}{10^{-3} cm^{-3}}\right) \left(\frac{R}{10 kpc}\right)^3 M_\odot \quad (2)$$

Gas masses (given in Table 4) range from $2.2 \times 10^8 - 1.5 \times 10^{10}$ Solar masses with an average of $3.0 \times 10^9$ Solar masses.

The total mass is calculated from the Chandrasekar-Emden equation for a self-gravitating isothermal sphere.

$$\frac{1}{\xi^2}\frac{d}{d\xi}\left(\frac{\xi dV}{d\xi}\right) = e^{-V} \text{ with } \xi = r/a \quad (3)$$

$$For\ \xi \rightarrow r \gg a, the\ total\ mass\ density, \rho \simeq \frac{kT}{2\pi G \mu m_H r^2}$$

The total mass density includes both visible matter and dark matter. Solving the mass integral,

$$M = \int_0^R 4\pi r^2 \rho dr \text{ gives the total gravitating mass: } M \simeq \frac{2kTR_x}{G\mu m_H} \quad (4)$$

The mass scales with source radius ($R_x$) and temperature. We calculate the masses out to the galactic radius, given in Table 1. The mean molecular weight, µ, comes into the calculation through the gas pressure. For primordial abundances, and assuming total ionization, µ, is 0.61. The scaled mass is:

$$M = 6.4 \times 10^{10} \left(\frac{T}{10^7 K}\right) \left(\frac{R}{10 kpc}\right) M_\odot \quad (5)$$

The total masses are given in Table 4 and range from $2.8 \times 10^{10} - 3.6 \times 10^{11}$ Solar masses with an average of $1.3 \times 10^{11}$.

## Table 4: Gas Density and Derived Masses

| Galaxy ID | Gas Density | Gas Mass | Total Mass | M/L Ratio | Gas Mass Fraction |
|---|---|---|---|---|---|
| | $10^{-3}$ cm$^{-3}$ | $10^9$ Solar Masses | $10^{11}$ Solar Masses | | |
| DXG1 | 3.7 | 15.3 | 3.6 | 2500 | 0.042 |
| DXG2 | 8.7 | 2.1 | 1.3 | 2660 | 0.017 |
| DXG3 | 4 | 13.7 | 3.1 | 2250 | 0.044 |
| DXG4 | 13 | 2.15 | 1.6 | 2410 | 0.008 |
| DXG5 | 8.5 | 0.51 | 1 | 9180 | 0.005 |
| DXG6 | 25 | 0.22 | 0.6 | 4240 | 0.004 |
| DXG7 | 21 | 1.07 | 0.8 | 1474 | 0.013 |
| DXG8 | 2.3 | 0.57 | - | - | |
| DXG9 | 6.9 | 0.67 | 1.1 | 9333 | 0.006 |
| DXG10 | 21 | 1.56 | 0.3 | 3333 | 0.056 |
| DXG11 | 6.6 | 0.95 | 1.2 | 8000 | 0.008 |
| DXG12 | 27 | 0.43 | 0.4 | 6307 | 0.01 |
| DXG13 | 4.6 | 4.04 | 1.2 | 2574 | 0.034 |
| DXG14 | 13 | 1.66 | 0.4 | 6545 | 0.046 |
| DXG15 | 6.4 | 0.79 | 1.2 | 9600 | 0.007 |
| DXG16 | 3.2 | 3.06 | 2.3 | 8654 | 0.014 |

## 4. Discussion

The characteristic temperature of the DXGs can be compared with that of the ICM. The best constraint on the temperature of the Abell 1367 ICM comes from modeling the broad, high quality (0.4 - 20 keV) spectrum from three combined X-ray data sets - the RXTE PCA, the ASCA GIS, and the ROSAT PSPS - with multiple spectral components (Henriksen & Mushotzky 2001). The best fitting model is two thermal components: 1.11 - 1.55 keV and 4.35 - 4.72 keV. The DXG temperature range overlaps the lower temperature component of the ICM. The integrated emission from DXGs could be part of the low temperature component but not all as the soft emission generally has 20 – 30% Solar abundance and the DXGs are significantly lower.

The range of hot gas mass fraction (= gas mass/total mass), 0.4 - 5.6% for DXGs, is overlapping of early type galaxies: 0.1 - 1.0% (Babyk et al. 2018) . However, the average total gas fractions, including: hot, HI, and H$_2$, for early-type galaxies is 0.57%. The averages are based on detected masses for 42 early-type galaxies from the ATLAS survey, which are detected in the X-ray (Su et al. 2015) . This value overlaps the low end of the DXG gas fraction range. But the average DXG hot gas mass is significantly higher, ~10x times, than the average early-type hot gas mass. The average gas mass fraction for DXGs, 2.3% is approaching that of Sa type galaxies, 4% (Young & Scoville 1991). The similarity between the HI dominated cold gas mass in Sa spirals and the Hydrogen dominated gas fraction of DXGs may hint at a process of gas heating and replacement in spiral galaxies.

Figure 1 shows that the X-ray sources span a distance of 40 arcmin or 1.2 Mpc. While they are not concentrated to the center of the cluster, they inhabit the inner 25% of the sampled region. This is for Z = 0.022 (Struble & Rood 1999) and $H_0$ = 69.6, $\Omega_m$ = 0.286, $\Omega_v$ = 0.714 (Wright 2006). This may indicate a formation process connected to cluster evolutionary mechanisms, all of which are stronger toward the center. But typical stripped gas components appear to be too small (discussed above) or in the case of halos, too large ($3.0 \times 10^9$ Solar masses for DXGs compared to $1.4 \times 10^{10}$ Solar Masses for coronae around early-type galaxies (Forman et al. 1985). Based on models for the formation of the galactic coronae (Forman, Jones & Tucker 1994), we conclude that DXGs are not consistent with stripped coronal gas.

The DXGs also have a dominant dark matter component, which argues for a dark galaxy interpretation over a strictly gas origin. On dynamical grounds, it seems unphysical that the X-ray sources are merely stripped gas offset from the nearest optical galaxies. The free-fall time, assuming the optical galaxy has a similar mass to the X-ray sources, ranges from 17 million to 135 million years. For the average separation of 17.6", the average free-fall time is 58 million years. This would be an unstable configuration with a low probability of observation. In fact, their observation at the present frequency would necessitate that there be hundreds of these systems, making them a dominant component of the total cluster mass.

In addition, there are vast differences in average gas mass, gas mass fraction, average gas temperature (0.31 keV) (Su et al. 2015), and metallicity for known early-type galaxies, known to be X-ray emitters. The metallicity difference is significant. Based on a sample of 32 early type galaxies, the average metal content of the hot gas component is 30% Solar and none have less than 10% (Su & Irwin 2013), while the average for DXGs is 2.6 – 7% (all values ae 90% confidence).

A detailed study of a recently stripped galaxy, UGC 6697 in the Abell 1367 cluster, is given in (Consolandi et. al. 2017). These authors compare the kinematics and distribution of both gas and stars, separating the ionized gas as traced by H$\alpha$, into both a bound component and a stripped component. Simulations by (Tonnesen & Bryan 2010) show that outer, low density gas is stripped, while higher density gas is retained by the galaxy. There is mixing of the hot ICM with the retained interstellar gas which can lead to heat, compression and cooling of the bound gas, followed by star formation(Kronberger et al. 2008). Cold gas may be accelerated in the resulting stellar wind, growing in mass, if the radiative cooling time of mixed gas is shorter than the cloud destruction time (Gronke & Oh 2018). This could lead to a short period of enhanced star formation followed by quenching of star formation, as the cold gas is lost to the ICM. The galaxy will then fade and become red as it passively ages. Alternatively, we suggest that such a galaxy may appear as a DXG rather than faded away in the optical. This hypothesis is supported by the detection of magnetic energy from a bow shock in the galactic outflow of M82. The extrapolated turbulent kinetic energy dominates the magnetic field energy at ~ 7 kpc. This implies that the magnetic field lines are 'open' to the surrounding ICM of the Virgo Cluster (Lopez-Rodriguez et al. 2021) . With an untangled magnetic field structure, heating by conduction from the surrounding ICM will proceed turning the inert galaxy into a reservoir of hot gas. The dark galaxies are on the low end of the mass spectrum of normal galaxies and their precursors may have been more easily stripped and quenched within the evolving cluster environment. Their evolution with the ICM fully truncates star formation leaving the galaxy barren of stars and filled with a hot plasma.

The ram-pressure stripping criteria (Sarazin 1988) is recast using the DXG observables to estimate the stripping radius ($R_s$). Inside of the $R_s$, the galaxy is not stripped. The values for the DXGs, $T_x$ and $R_x$ are the composite hot gas temperature and average DXG radius, respectively, and are related to the total mass of the galaxy. The line-of-sight velocity dispersion is for the inner region of the cluster (Dickens & Moss 1976) , the ICM density is a nominal value for the inner region of a rich cluster, and the cold gas fraction is typical of Sa type spiral galaxies (Young & Scoville 1991).

$$\left(\frac{R_s}{14 kpc}\right) = \left[\left(\frac{T_x}{1.35 keV}\right)^2 \left(\frac{\overline{R_x}}{12.7 kpc}\right)^2 \left(\frac{f_g}{0.04}\right)\left(\frac{10^{-3}}{n_{ICM}}\right)\left(\frac{700 km/s}{V_{los}}\right)^2\right]^{1/4}$$

There is pretty good agreement between the estimated size of the stripped disk and the average DXG radius. This is consistent with scenario outlined in this paper that identifies stripped spiral galaxies as the precursors of DXGs.